\documentclass[preprint,superscriptaddress,nofootinbib,showkeys,showpacs,
tightenlines,fleqn]{revtex4} 
\usepackage{graphicx}
\newcommand{\be}{\begin{equation}}
\newcommand{\ee}{\end{equation}}
\newcommand{\bee}{\begin{eqnarray}}
\newcommand{\eee}{\end{eqnarray}}
\newcommand{\eq}{\end{quote}}
\newcommand{\nn}{\nonumber}
\begin{document}
\preprint{PNU-NTG-12/2005}
\preprint{PNU-NURI-3/2006}
\title{The leading-twist pion and kaon distribution amplitudes from 
the QCD instanton vacuum} 
\author{Seung-il Nam}
\email{sinam@pusan.ac.kr}
\affiliation{Department of Physics and Nuclear physics \& Radiation Technology
  Institute 
(NuRI), Pusan National University, Busan 609-735, Korea}
\author{Hyun-Chul Kim}
\email{hchkim@pusan.ac.kr}
\affiliation{Department of Physics and Nuclear physics \& Radiation Technology
  Institute 
(NuRI), Pusan National University, Busan 609-735, Korea}
\author{Atsushi Hosaka}
\email{hosaka@rcnp.osaka-u.ac.jp}
\affiliation{Research Center for Nuclear Physics (RCNP), Ibaraki, Osaka
  567-0047, Japan} 
\author{M.~M.~Musakhanov}
\email{musakhanov@pusan.ac.kr}
\affiliation{Theoretical Physics Department, National Univeristy of
  Uzbekistan, Tashkent 7000174, Uzbekistan}  
\date{\today}
\begin{abstract}
We investigate the leading-twist light-cone distribution
amplitudes for the pion and kaon, based on the nonlocal chiral quark
model from the instanton vacuum.  Effects of explicit flavor SU(3)-symmetry
breaking are taken into account.  The Gegenbauer moments are  
computed, analyzed and compared with those of other models.  The
one-loop QCD evolution of the moments is briefly discussed.  The
transverse momentum distributions are also discussed for the
pion and kaon light-cone wave functions.
\end{abstract}
\pacs{12.38.Lg, 14.40.Ag}
\keywords{Meson Distribution Amplitudes, Flavor SU(3)-Symmetry
  Breaking, Instanton}
\maketitle
\section{Introduction}
The meson light-cone distribution amplitude (DA) is of great
importance to understand the key properties of mesons in exclusive
hardronic processes, since it provides essential information on the 
non-perturbative structure of a meson in the
processes~\cite{Efremov:1979qk,Lepage:1979zb,Lepage:1980fj,Chernyak:1983ej}.
An experimental aspect of the pion DA can be seen in the recent
measurements of $\pi$--photon transition form factor by the CLEO
experiment~\cite{Gronberg:1997fj}. The analysis of these data in
Ref.~\cite{Schmedding:1999ap} has shown that neither two-humped DA for
the pion predicted by Chernyak and Zhitnitsky
(CZ)~\cite{Chernyak:1977fk} nor the asymptotic one are favored at the
$2\sigma$ level of accuracy.  These findings were independently
confirmed by Bakulev \textit{et al.} in
Ref.~\cite{Bakulev:2002uc,Bakulev:2003cs}.  On the theoretical side, 
the pion DA has been investigated in various 
approaches: For example, in the QCD sum
rules
(QCDSR)~\cite{Chernyak:1983ej,Braun:1988qv,Bakulev:1991ps,Bakulev:1994su,
Bakulev:2001pa,Bakulev:2005vw},  
in lattice QCD (LQCD)~\cite{DelDebbio:2002mq,Dalley:2002nj}, in the chiral
quark model from the instanton
vacuum~\cite{Petrov:1998kg,Dorokhov:2002iu,Dorokhov:1991nj,
Dorokhov:2002tq,Anikin:1999cx,Dorokhov:2001ys,Esaibegian:1989uj},
in the NJL models~\cite{Praszalowicz:2001wy,Praszalowicz:2001pi,
RuizArriola:2002bp}, in chiral perturbation theory
($\chi$PT)~\cite{Chen:2005js} and so on.  A comparison of the results
from those models with the CLEO data was lucidly made in
Ref.~\cite{Bakulev:2005cp}, all the results being evaluated at
$\mu^2=5.76\,{\rm GeV}^2$ (known as the Schmedding-Yakovlev (SY)
scale~\cite{Schmedding:1999ap}) after NLO evolution.   

While the pion DA has been extensively studied for well over decades,
the kaon one has attracted attention rather recently, since it is of
deep relevance to the exclusive $B$-meson decays to light mesons.  The
$B$-meson decays are now being under full investigation at the BaBar
and Belle experiments.  These experiments will soon shed light on the
pattern of the $CP$-violation as well as of flavor SU(3)-symmetry
breaking.  In this context, there has been some amount of theoretical
works on the DA in the QCDSR~\cite{Ball:2003sc,Braun:2004vf, 
Khodjamirian:2004ga,Ball:2005ei,Ball:2005vx}.  Though CZ derived
kaon DA, originally in the QCDSR~\cite{Chernyak:1981jd,
Chernyak:1982it}, their results are known to suffer from the sign
mistake.   

It is known that spontaneous chiral symmetry breaking is
well realized in the instanton vacuum via quark zero modes.  Thus, it
may provide a good framework to study the structure of light mesons.
Moreover, there are only two parameters in this approach: The average
instanton size $\rho\approx \frac{1}{3}$ fm and average
inter-instanton distance $R\approx 1\, {\rm fm}$.  In particular, the
normalization scale of the present approach is naturally identified by
the average size of instantons and is approximately equal to
$\rho^{-1}\approx 600$ MeV.  The values of the $\rho$ and $R$ were
estimated phenomenologically by Ref.~\cite{Shuryak:1981ff} as well as   
theoretically by Refs.~\cite{Diakonov:1983hh,Diakonov:1985eg}.  These
values were recently confirmed in various lattice simulations of the
QCD vacuum \cite{Chu:vi,Negele:1998ev,DeGrand:2001tm}.  Very recent
lattice calculations of the quark propagator~\cite{Faccioli:2003qz,
Bowman:2004xi} are in a remarkable agreement with that of
Ref.~\cite{Diakonov:1983hh}.  The instanton vacuum model was later
extended by introducing the current-quark
masses~\cite{Musakhanov:1998wp,Musakhanov:2001pc,Musakhanov:vu}.  It 
was assumed in the model that the large $N_c$ expansion is the
reasonable one and the results were obtained in the leading order in
this expansion.  In the present work, we want to investigate the 
leading-twist pion and kaon DAs within the framework of the nonlocal
chiral quark model ($\chi$QM) from the instanton vacuum, with SU(3)-symmetry
breaking effects taken into account~\cite{Musakhanov:1998wp, 
Musakhanov:2001pc,Musakhanov:vu}.  We basically follow the formalism
in Ref.~\cite{Praszalowicz:2001wy} apart from flavor SU(3)-symmetry breaking
effects.    

Although the derivation of the effective action with flavor SU(3)-symmetry
breaking was rather complicated, the final result is 
summarized in the denominator of the quark propagator with explicit flavor
SU(3)-symmetry 
breaking in a very simple form, $\rlap{/}{k}+i[m_f+M_f(k)]$.  The only 
difference is the explicit appearance of the 
current-quark mass in the denominator when it is compared to the
propagator in the chiral limit ($m_s=m_u=0$).  We note here
that $M_f(k)$ is the dynamical quark mass depending on the quark
momentum and current-quark mass.  However, instead of using the
non-local quark mass derived from the instanton vacuum, we employ in
the present work a simple-pole parameterization of the momentum-dependent
quark mass $M_f(k)$, since it shows a similar behavior to the original
one and it can be easily continued analytically to Minkowski space.

The $\chi$QM from the instanton vacuum has one great virtue: The
normalization point is naturally given by the average instanton
size whose value is determined by the saddle-point equation.  However,
we will fix the scale by the normalization condition for the
DAs in the present work in such a way that we can reproduce the pion
and kaon decay constants simultaneously.  Employing 
the dipole-type form factor, we get the cut-off mass $\Lambda$ to be
around 1.2 GeV which plays a role of the scale of the present method.
However, since the evolution does not depend sensitively on the scale,
we will take our scale to be 1 GeV which is not far from the original
scale~\cite{Petrov:1998kg}.  From this value, we can evolve
the DAs obtained in this approach to the scale of the CLEO experiment,
$\Lambda\simeq 2.4\,{\rm GeV}$ (the SY scale~\cite{Schmedding:1999ap})
so that we can compare the results with the empirical data.  

We will provide in the present work the following numerical results:
The DAs, the Gegenbauer moments, the moments of the amplitudes, the
transverse momentum distributions and their ratios for the pion and
kaon.  We observe the symmetric and asymmetric shapes for the pion and
kaon, respectively, as expected from the effects of explicit flavor
SU(3)-symmetry breaking.  However, when we take into  
account the current-quark mass corrections to the dynamical quark
mass, the kaon DA becomes less asymmetric.  We also
consider the one-loop QCD evolution of the DA so that we may compare
our results with those from the other models.  Our results are 
in good agreement with others qualitatively.  The moments of
the distribution amplitudes, $\langle\xi^m\rangle$, are also given for the
pion and 
kaon.  Finally, we consider the transverse momentum distributions of
the full light-cone wave function, $\Psi(u,k_T)$.  Using this, we test
the factorization hypothesis for the kaon.  

This paper is organized as follows: In Section II, we briefly explain
the general formalism to derive the DAs within the framework of the
nonlocal $\chi$QM from the instanton vacuum.  The numerical results
are given in Section III with discussions.  The last Section is
devoted to summary and conclusion.

\section{General formalism}
In the present Section, we briefly explain how to derive the pion and
kaon light-cone DAs within the framework of the nonlocal $\chi$QM from
the instanton vacuum.  The leading-twist (twist-two) light-cone DA for
the pion and kaon is defined as follows~\cite{Chernyak:1977fk}:
 \begin{eqnarray}
\Phi_{\phi}(u)
=\frac{1}{i\sqrt{2}F_{\phi}}\int^{\infty}_{-\infty}\frac{dz}{2\pi}e^{-i(2u-1)
z\cdot P}\langle0|\bar{q}_{f}(z)\rlap{/}{\hat{n}}\gamma_{5}\exp\left[
ig\int^{z}_{-z}dz'^{\mu}A_{\mu}(z^{\prime})\right]q_{g}(-z)|\phi(P)\rangle,    
\label{DAAV}
\end{eqnarray}
where $|\phi(P)\rangle$ is the state vector for a pesudoscalar meson
moving with on mass-shell momentum $P$ in the light-cone
frame.  $\psi$ and $A_{\mu}$ stand for the quark and gluon fields,
respectively.  $u$ and $z$ indicate the longitudinal momentum fraction
and spatial separation between the quarks. $n_{\mu}$ is the light-like
vector satisfying the condition of $n^2=0$. The exponential term is
called the Willson line or gauge connection which guarantees the
gauge invariance of the nonlocal quark bilinear operator.  However, by
virtue of the light-cone gauge, we set this Wilson line to be
unity.  Therefore, we need not consider any gluon contribution for
leading-twist pion and kaon DAs.  The pseudoscalar (PS) meson decay
constant $F_{\phi}$ is introduced as the normalization
constant, which will be determined with the cut-off mass $\Lambda$ in
the model.  The DA must fulfill the following normalization condition:   
\begin{eqnarray} 
\int^{1}_0du\,\Phi_{\phi}(u)=1.  
\label{normalization}
\end{eqnarray}
We will fix the cut-off mass $\Lambda$ which is approximately
identified as the scale of the present approach, using
Eq.~(\ref{normalization}).  In order to calculate the pion and kaon
DAs, we first start with the partition function of the nonlocal
$\chi$QM in Euclidean space:   
\begin{eqnarray} 
\mathcal{Z}&=& \int {\cal D} \psi {\cal D} \psi^\dagger
{\cal D} \phi^a
\exp \int d^4 x \Big[ \psi^{\dagger \alpha}_{f} (x)
(i \rlap{/}{\partial} +i m_f)\psi^{\alpha}_{f}(x)\nn\\
&+&i\int \frac{d^4 k d^4 l}{(2\pi)^8} e^{i(k-l)\cdot x}
\sqrt{M_f(k) M_g(l)} \psi^{\dagger \alpha}_{f} (k)
\left(U^{\gamma_5}\right)_{fg}
\psi^{\alpha}_{g} (l)\Big].
\label{Lagrangian}
\end{eqnarray}
This partition function can be constructed by considering arbitrary $N_f$,
large $N_c$ and the saddle-point equation based on the instanton
induced $2N_f$-quark interaction.  The subscripts $f$ and $g$ denote
the quark-flavor indices. In the present work, we have
$(f,g,\phi)=(s,u,K^+)$ and $(d,u,{\pi}^+)$. $\alpha$ 
represents the color index.  Note that the quark and anti-quark
interact each other nonlocally via the nonlinear PS-background field
of $\phi^a$. The dynamical quark mass $M_f(k)$ plays a role of the 
quark and PS-meson coupling strength, which arises originally 
from the Fourier transform of the quark zero-mode solutions.
Moreover, this momentum-dependent quark mass can be regarded as a natural 
and intrinsic UV regulator for loops in the model.  The nonlinear
background PS-meson field is defined as follows: 
\begin{equation}
U^{\gamma_5}=U(x) \frac{1+\gamma_5}{2} + U^\dagger (x)
\frac{1-\gamma_5}{2} \;=\;
\exp\left({i\gamma_5\phi^a(x) \lambda^a}/{F_{\phi}} \right),  
\label{ufield}
\end{equation}
where $\lambda^a$ is the flavor SU(3) Gell-Mann matrices.   the quark
propagator can be easily obtained from the effective chiral action
given in Eq.~(\ref{Lagrangian}).   One of the authors (M.M.) suggested
the effective chiral action in terms of the multi-quark interaction
with explicit flavor SU(3)-symmetry breaking~\cite{Musakhanov:1998wp}.
This action is called the modified improved action (MIA), which can be
distinguished from the usual effective action derived by Diakonov and
Petrov~\cite{Diakonov:1983hh}.  The quark propagator from the MIA is
obtained as follows:
\begin{eqnarray}
\frac{1}{D}_f=\frac{i}{\rlap{/}{k}+i[m_{f}+M_f(k)]}.
\label{propagator}
\end{eqnarray}
Note that the denominator contains the current-quark mass $m_f$
explicitly.  Moreover, the dynamical quark mass $M_f(k)$ also depends
on the $m_f$.  We parameterize in the present work the dynamical quark
mass as follows:  
\begin{eqnarray}
M_f(k)=M_{0}F^2(k^2)f(m_f)=M_{0}\left[\frac{n\Lambda^2}{(n\Lambda^2+k^2)}
\right]^{2n}\left[\sqrt{1+\frac{m^2_f}{d^2}}-\frac{m_f}{d}\right].
\label{dqm}
\end{eqnarray}
This parameterization of a simple-pole type resembles the behavior
of the Fourier transform of the quark zero modes.  $M_0$ is the
constituent quark mass being set to be $350$ MeV.  

We consider the power $n$ in Eq.~(\ref{dqm}) as a free
parameter~\cite{Praszalowicz:2001wy}.  The form of the $m_f$ 
correction factor $f(m_f)$ in Eq.~(\ref{dqm}) was suggested by
Ref.~\cite{Musakhanov:1998wp} via the saddle-point equation of the
effective action.  The explicit expression of $f(m_f)$ was   
derived by Pobylitsa: If one expands quark propagator in the
presence of the instanton effect and resums quark loops which are  
suppressed in the large $N_c$ expansion, one gets the correction 
factor $f(m_f)$ as given in Eq.~(\ref{dqm})~\cite{Pobylitsa:1989uq}.
For convenience, we denote the effective action without the correction
factor by {\it MIA1} and that with it by {\it MIA2}. 

It is worth mentioning that there is a caveat in the present
approach.  The effective low-energy QCD partition function of
Eq.~(\ref{Lagrangian}) is well defined in Euclidean space.  However, we
have to continue analytically to Minkowski space, in order to study   
the light-cone DAs.  Though we have no theoretically firm grounds for
such analytic continuation, we want to adopt a more practical stance on
it. Since the nonlocal $\chi$QM from the instanton vacuum was used for
investigating the light-cone DAs with this analytic continuation
successfully~\cite{Petrov:1998kg}, we will proceed to study the 
DAs in Minkowski space very carefully.  

The DAs in the present approach is obtained as follows: 
\begin{eqnarray} 
\Phi_{\phi}(u)&=&i\frac{N_c}{F^2_{\phi}}\int\frac{d^4k}{(2\pi)^4}
\delta(2uP\cdot \hat{n}-2k\cdot \hat{n})  \nonumber \\
&\times&\mathrm{tr}\left[\frac{\sqrt{M_f(k)}}{\rlap{/}{k}
-[m_f+M_f(k)]}\rlap{/}\hat{n}\gamma_5\frac{\sqrt{M_f(k-P)
}}{(\rlap{/}{k}-\rlap{/}{P})-[m_g+M_g(k-P)]}\gamma_5\right]  \nonumber \\
&=&i\frac{N_cM_0n^{2n}\Lambda^{4\mathrm{n}}}{F^2_{\phi}}\int\frac{d^4k}{
(2\pi)^4}\delta(2u P\cdot \hat{n}-2k\cdot \hat{n})\mathrm{tr}\frac{
[F_f(\rlap{/}{k}-\rlap{/}{P})+G_{f}]\rlap{/}{n}(F_g\rlap{/}{k}+G_{g})}{D_fD_g
},  \nonumber \\
\label{DAAV1}
\end{eqnarray}
where the functions $F,\, G$ and $D$ are given as follows:
\begin{eqnarray}
F_{f}&=&(k^2-n\Lambda^2)^{3\mathrm{n}},  \nonumber \\
G_{f}&=&(k^2-n\Lambda^2)^{\mathrm{n}}[m_{f}(k^2-\Lambda^2)^{2\mathrm{n}
}+\eta],  \nonumber \\
D_{f}&=&(k^2-n\Lambda^2)^{4\mathrm{n}}(k^2-m^{2}_{f})-2m_{f}\eta(k^2-n
\Lambda^2)^{2\mathrm{n}}-\eta^2+i\epsilon,  \nonumber \\
F_{g}&=&[(k-P)^2-n\Lambda^2]^{3\mathrm{n}},  \nonumber \\
G_{g}&=&[(k-P)^2-n\Lambda^2]^{\mathrm{n}}[-m_{g}[(k-P)^2-n\Lambda^2]^{2
\mathrm{n}}-\eta],  \nonumber \\
D_{g}&=&[(k-P)^2-n\Lambda^2]^{4\mathrm{n}}[(k-P)^2-m^{2}_{g}]-2m_{g}\eta
[(k-P)^2-n\Lambda^2]^{2\mathrm{n}}-\eta^2+i\epsilon  \nonumber
\end{eqnarray}
with $\eta=M_0\Lambda^{4n}$.  We evaluate Eq.~(\ref{DAAV1}) to order  
$\mathcal{O}(m^2_{\pi,K})$, using the relations
$P^2=P^+P^--P^2_T\simeq P^+P^-=m^2_{\pi,K}$ in the light-cone 
frame.  We set $m_{\pi}=140$ MeV and $m_K=495$ MeV for numerical input.   
\section{Numerical results}
First, we show briefly how to fix the cut-off mass $\Lambda$.  We set 
$\Lambda$ by using the normalization condition given in
Eq.~(\ref{normalization}) in such a way that it reproduces the
empirical $\pi$ and $K$ meson decay constants simultaneously.  We
first consider the case of $n=1$.  In this case, we obtain
$\Lambda=1.2\, {\rm GeV}$ with which the pion decay constant
$F_{\pi}=96.77$ MeV ($93$ MeV) and the kaon decay constant
$F_K=110.21$ MeV ($113$ MeV) are well reproduced qualitatively in
comparison with the experimental data put in the parentheses.  We use
the value of $\Lambda=1.2$ GeV for all cases of $n$ and list the
results of the calculated pion and kaon decay constants in
Table~\ref{table1}.  As shown in Table~\ref{table1}, we find that the
values of the pion and kaon decay constants decrease as $n$ increases.
In the case of the MIA1, they are qualitatively in a good agreement
with the data, while for the MIA2 the kaon decay constants are
underestimated by about $20\,\%$.  It can be understood as follows:
The current-quark mass corrections reduce $M_f(k,m_f=0)$ by about
$25\%$ for the strange quark whereas they lessen it by about $2\%$ for
the up and down quarks.  We will see later that the kaon distribution
amplitudes is also reduced similarly with the MIA2 used.
\begin{table}[ht]
\begin{tabular}{c|ccc|ccc}
\hline
&\multicolumn{3}{c|}{$F_{\pi}$}&\multicolumn{3}{c}{$F_{K}$}\\
\hline
$n$ & $1$ & $2$ & $3$ & $1$ & $2$ & $3$\\
\hline
MIA1 & 96.77 & 93.66 & 92.83 & 110.21 & 107.22 & 106.36\\
MIA2 & 95.13 & 92.45 & 91.76 & 98.51 & 96.47 & 96.07\\
\hline
\end{tabular}
\caption{The results of the pion decay constant $F_{\pi}$ and the kaon
decay constant $F_K$ in units of MeV with $\Lambda=1.2$ GeV.}
\label{table1}
\end{table}
Though we are able to fit the kaon decay constant for the MIA2 by
changing the cut-off mass, we will not try to do that, since we want 
to produce all results with one cut-off mass.  Moreover, 
note that the meson-loop corrections {($\sim1/N_c$)}
are not taken into account in the present work, which is known to be
of great significance to reproduce the proper value of 
the kaon decay constant as shown in $\chi$PT as well 
as in the   nonlocal $\chi$QM
calculations~\cite{Gasser:1984gg,Kim:2005jc,Kimetal}.  However, we do 
not take into account any $1/N_c$ correction in the present work,
since it is beyond the aim of the present work.  Thus, we still 
have room to improve the result of the kaon decay constant in the
future~\cite{Kimetal}, including the meson-loop corrections. 

The presence of the nonlocal interaction between quarks and Nambu-Goldstone
bosons, which arises from the momentum-dependent quark mass, breaks
the gauge invariance, so that the currents are not conserved.  The
conserved currents in Euclidean space with the nonlocal interactions
can be derived by gauging the partition
function~\cite{Musakhanov:2002xa}.  The correct expression of the pion
decay constant $f_\pi^2$ can be derived by using 
the modified axial-vector current in the following matrix elements:
\begin{equation}
\left<0\left|A_\mu^a (x)\right| \pi^b(P)\right> = i f_\pi P_\mu
e^{iP\cdot x} \delta^{ab},
\end{equation}
which indicates that the Ward-Takahashi identity of PCAC is well
satisfied with the modified conserved axial-vector current.  If we use
the usual form of the axial-vector current $A_\mu^a =
\bar{\psi}\gamma_\mu\gamma_5\lambda^a\psi$, we would end up with the
Pagels-Stokar (PS) expression for $f^2_{\pi,\rm PS}$:
\begin{equation}
f^2_{\pi,\rm PS} = 4N_c\int \frac{d^4 k}{(2\pi)^4} \frac{M^2-\frac14
  MM' k}{(k^2+M^2)^2}\,\,\,\,{\rm (in\,\,Euclidean\,\,space)}.
\label{eq:ps}
\end{equation}
which gives smaller results than the correct expression by
approximately $20\,\%$.  Thus, one has to consider the modified
conserved currents in Eq.~(\ref{DAAV1}).  However, if we use the
$f^2_{\pi,\rm PS}$ for the normalization of the pion and kaon DAs for
convenience, we need not take into account the current conservation in
Eq.~(\ref{DAAV1}), since the results are almost the
same~\cite{NamKim}.   

\begin{figure}[ht]
\begin{tabular}{cc}
\includegraphics[width=8cm]{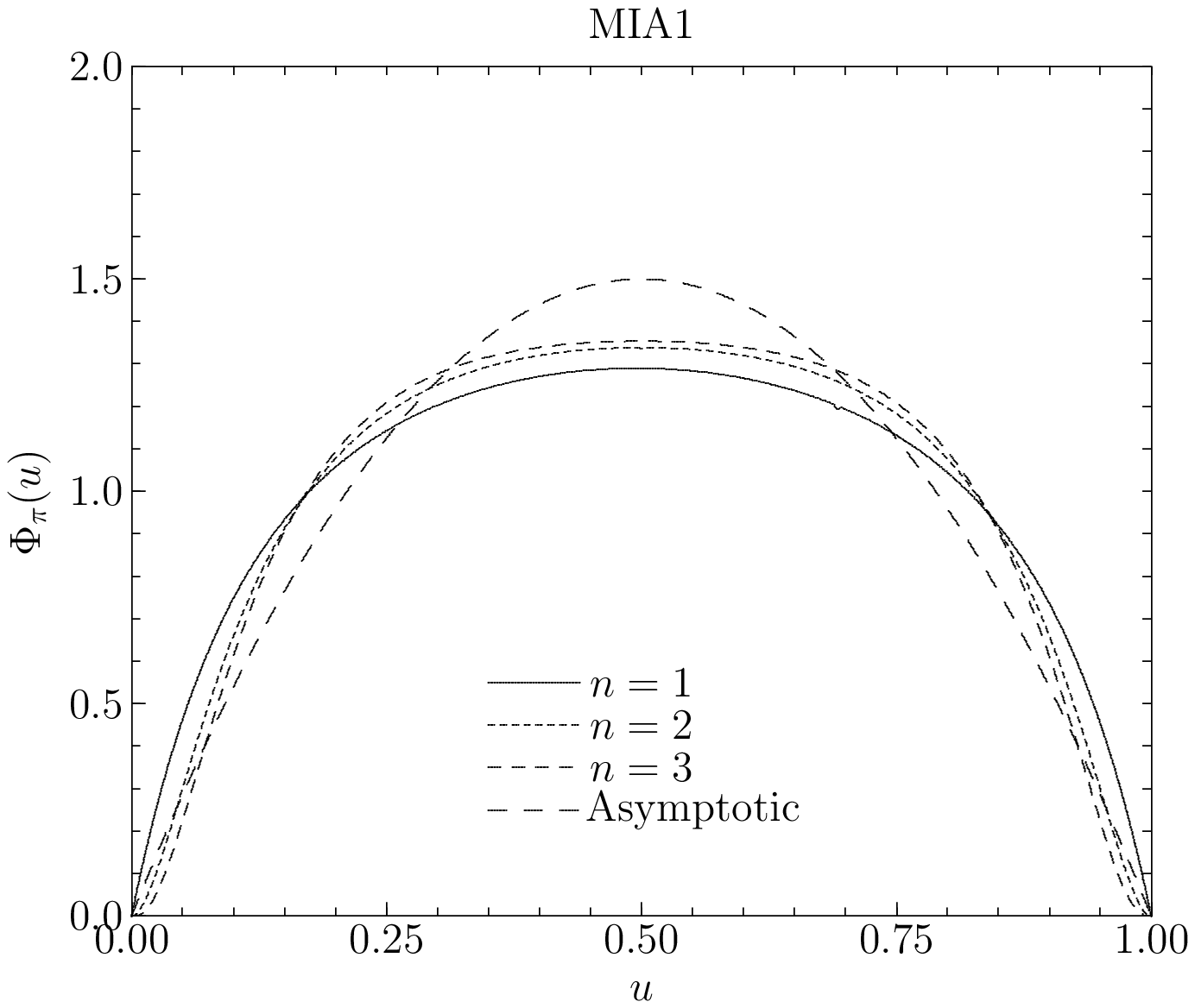}
\includegraphics[width=8cm]{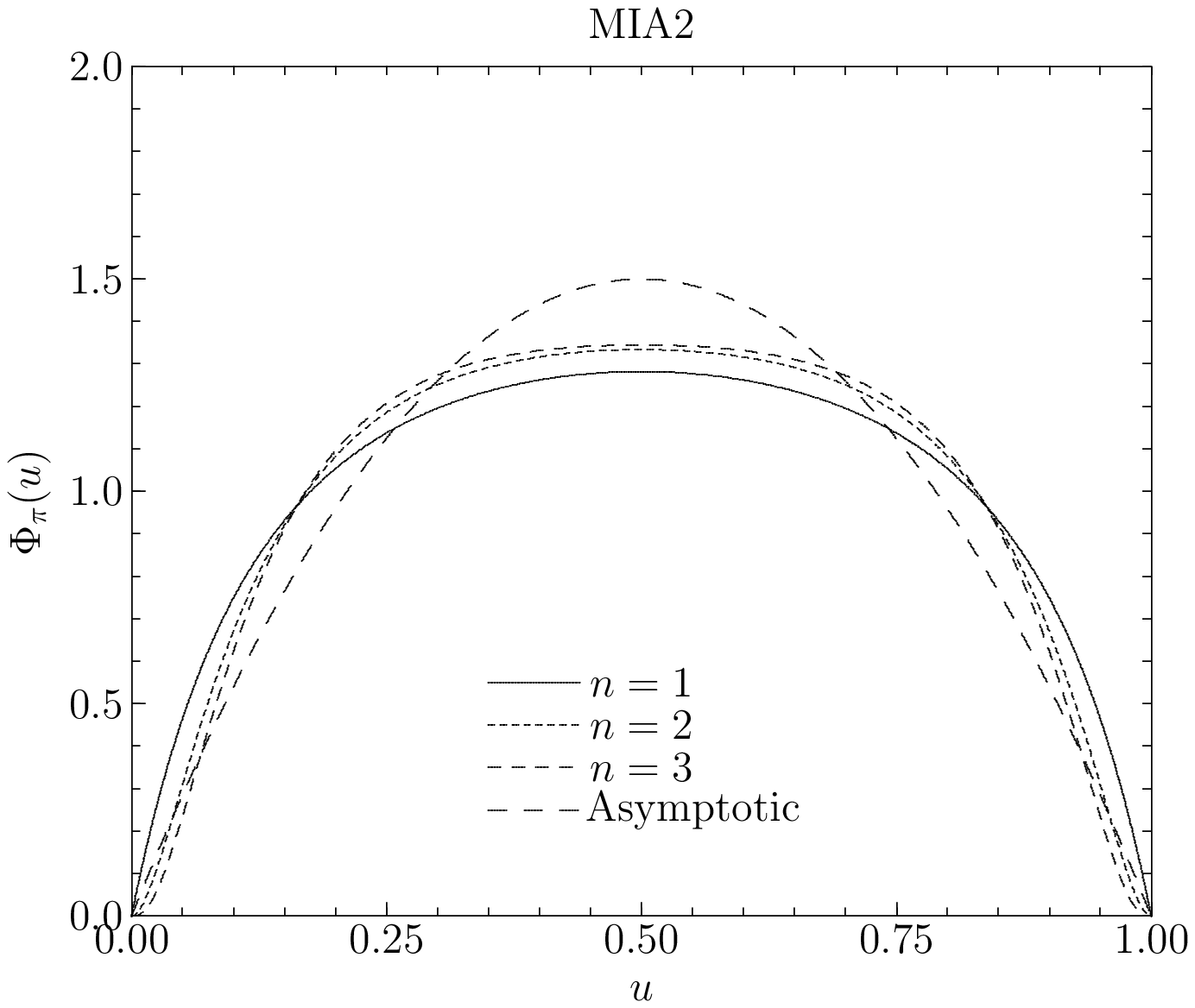}
\end{tabular}
\begin{tabular}{cc}
\includegraphics[width=8cm]{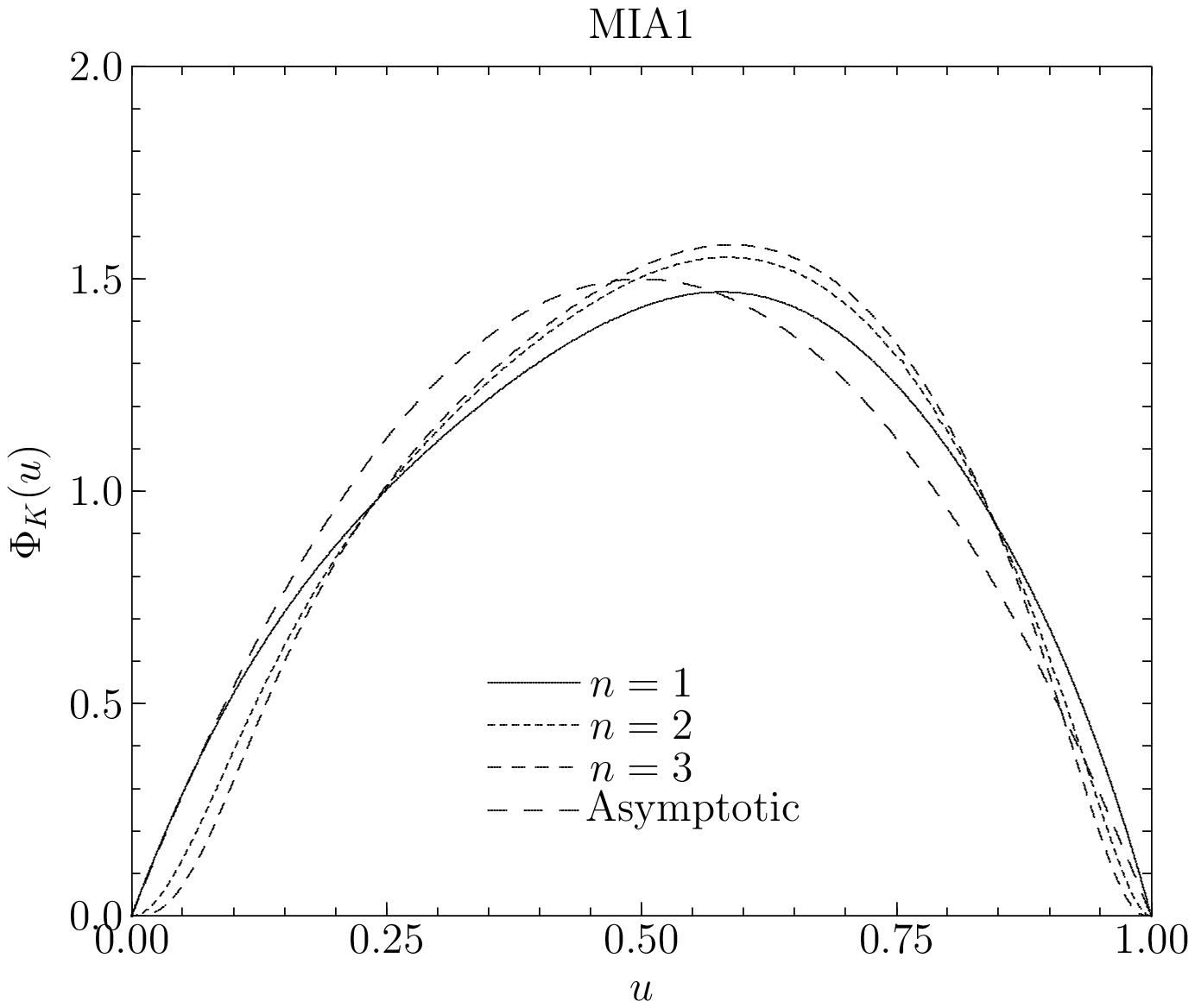}
\includegraphics[width=8cm]{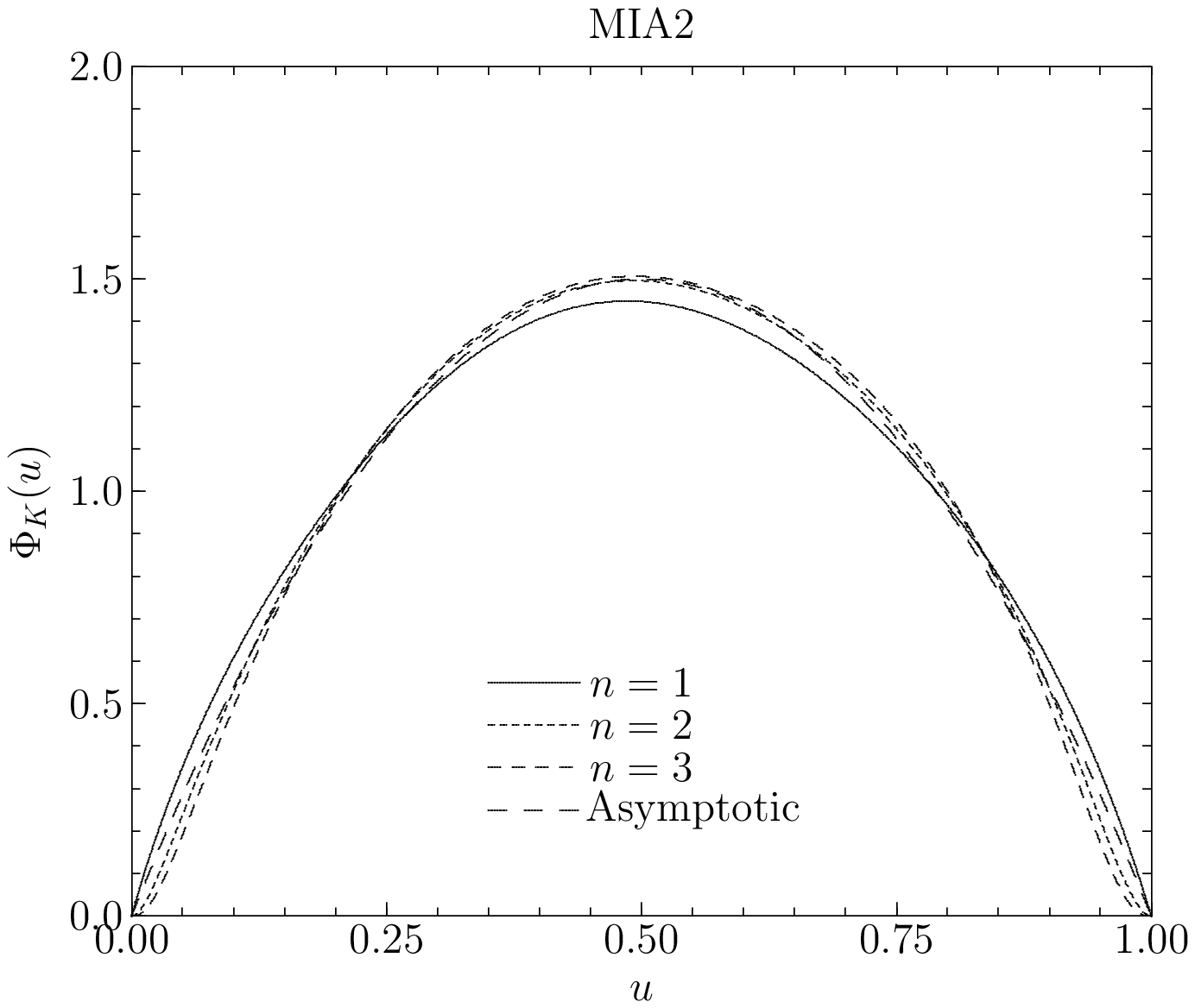}
\end{tabular}
\caption{DAs for the pion (upper two panels)
and kaon (lower two panels) with different powers 
$n$.  In the right column, we draw the distribution amplitudes for the 
MIA1, while we depict in the left column those for the MIA2.  We also
draw the asymptotic distribution amplitude in each panel for comparison.}   
\label{fig1}
\end{figure}

In Fig.~\ref{fig1} we depict the pion and kaon DAs, varying the power
$n$ with two different modified improved actions, MIA1 and MIA2.  The
asymptotic form of the distribution amplitude $\Phi_{\rm Asym}=6u(1-u)$
is plotted in each panel of the figure for comparison.  
The results of the pion DA are all symmetric under the inversion
$u\to1-u$ due to the fact that the mass of the 
quarks inside the pion are negligibly small ($m_u/\Lambda\sim
m_d/\Lambda\sim0$).  Thus, the present results are almost the same as those
in Ref.~\cite{Praszalowicz:2001wy} (only $2\,\%$ difference).  Note
that the current-quark masses for all flavors are taken into account
in the present work.  All curves are rather flat in the vicinity of
the centered area ($0.25\lesssim u\lesssim 0.75$),  
compared to the asymptotic one as already observed in
Ref.~\cite{Bakulev:2001pa}.  The pion DAs vanish at the end-points
$u=0$ and $1$, which indicates that it is impossible for a quark in the
pion to have zero or full momentum of $P$.  In the present approach,
this behavior is caused mainly by the term with $\sqrt{M_f(k)M_g(k)}$ in
the DA in Eq.~(\ref{DAAV1}).  It infers that the nonlocal interaction 
between quarks and pseudoscalar mesons is essential in describing the
correct end-point behavior of the DAs.    

Interestingly, as the power $n$ increases, the results of the pion DA behave
differently at the end-points, $u=0$ and $1$.  The results are getting
suppressed as the $n$ gets larger, so that the end-point behavior of
the DA changes from a convex form to a concave one.  This distinctive
behavior is of great importance for comparison with the empirical  
data~\cite{Schmedding:1999ap,Bakulev:2005cp}~\footnote{See
Refs.~\cite{Stefanis:1998dg,Stefanis:2000vd} for some relevant
discussion and for the role of the dynamical quark mass for the
end-point behavior.}.  

The kaon DAs are drawn in the lower panels of Fig.~\ref{fig1} for the MIA1
and MIA2.  The overall shapes and behavior at the end-points are very
similar to those of the pion ones.  However, as 
expected from explicit flavor SU(3)-symmetry breaking, the results
become asymmetric, since the quarks inside the kaon carry 
different fractions of the total momentum $P$ due to the mass
difference between the strange quark ($m_s=150$ MeV) and the up and
down quarks ($m_{u,d}=5$ MeV).  We find that
there is a noticeable difference between the cases of the 
MIA1 and MIA2.  As for the MIA1, the results look more asymmetric,
{\it i.e.} the right side is larger than the left side whereas those for
the MIA2 are almost symmetric.  This can be understood by recalling
the fact that the current-quark mass correction factor $f(m_f)$
reduces the dynamical quark mass by $\sim25\%$, so that the mass
difference between the light quark and strange quark gets smaller: 
\bee
\frac{m_u+M_u(k=0)}{m_s+M_s(k=0)}_{\rm MIA1}=0.71\hspace{0.5cm}
\Longrightarrow\hspace{0.5cm}\frac{m_u+M_u(k=0)}{m_s+M_s(k=0)}_{\rm
  MIA2}=0.84. 
\eee
Thus, the DAs for the MIA2 turn out to be nearly symmetric.  We note
that the results of the kaon DA for the MIA1 are rather similar 
to those calculated without the tensor term $\sim
k^q_{\mu}k^{\bar{q}}_{\nu}$ in Ref.~\cite{Szczepaniak:1993uq}. 
   
We are now in a position to study the Gegenbauer moments
for the pion and kaon DAs.  The Gegenbauer moments
are very useful in analyzing the DAs.  It can be
extracted from the experiments or derived from theoretical models.
In principle, the Gegenbauer moments indicate how much the
DAs deviate from the asymptotic one.  They are defined by the
coefficient of the expansion of the PS-meson DA in terms of the
Gegenbauer polynomials:   
\begin{eqnarray}
\Phi(u)&=& \Phi_{\rm Asym}(u)\sum^{
\infty}_{m=1}a_mC^{3/2}_m(\xi)  \nonumber \\
&=&6u(1-u)[a_0C^{3/2}_0(\xi)+a_1C^{3/2}_1(\xi)+a_2C^{3/2}_2(\xi)+\cdots],
\end{eqnarray}
where $a_0C^{3/2}_0(\xi)=1$ and $\xi=2u-1$. The coefficients $a_m$ are
called the Gegenbauer moments and can be calculated by using the
orthogonal condition for the Gegenbauer polynomials: 
\begin{equation}
\int^{1}_{-1} (1-\xi^2)C^{3/2}_m(\xi)C^{3/2}_{m^{\prime }}(\xi)=\frac{
\Gamma(m+3)}{m!(m+3/2)}\delta_{mm^{\prime }}.
\end{equation}
In Table~\ref{table2}, we list the calculated Gegenbauer moments for the
pion DAs with $n$ varied for the MIA1 and MIA2.
\begin{table}[ht]
\begin{center}
\begin{tabular}{c|ccc} \hline
$n$ & $a^{\pi}_2$ & $a^{\pi}_4$ & $a^{\pi}_6$\\ 
\hline
1, MIA1& $0.11012$ & $0.01500$  & $0.00153$\\
2, MIA1& $0.05330$ & $-0.02831$ & $-0.01597$\\
3, MIA1& $0.02903$ & $-0.04603$ & $-0.01897$\\
\hline
1, MIA2& $0.11399$ & $0.01583$  & $0.00188$\\
2, MIA2& $0.05963$ & $-0.02822$ & $-0.01523$\\
3, MIA2& $0.03442$ & $-0.04479$ & $-0.01917$\\
\hline
\cite{Chernyak:1983ej} & $0.56$ &--&--  \\
\cite{Schmedding:1999ap} (2.4 GeV) & $0.12\pm0.03$ &--&--\\ 
\cite{RuizArriola:2002bp} (2.4 GeV)&--& $0.044\pm0.016$
&$0.023\pm0.010$\\ 
\cite{Khodjamirian:2004ga} (1.0 GeV)& $0.26^{+0.21}_{-0.09}$ &--&--\\ 
\cite{Gockeler:2005jz} (2.24 GeV)& $0.236(82)$ &--&--\\
\cite{Ball:2006wn} (1.0 GeV)& $0.25\pm0.15$ &--&-- \\
\cite{Braun:1989iv} (1.0 GeV)& $0.44$  & $0.25$ &-- \\ \hline
\end{tabular}
\end{center}
\caption{The Gegenbauer moments for the pion distribution
amplitudes.  The numbers in the parentheses stand for the scales of
the corresponding works.}   
\label{table2}
\end{table}
Due to isospin symmetry, the odd Gegenbauer moments of the pion DA
become zero.  Therefore, we only need to consider the even 
moments. In spite of the fact that $u$- and $d$-quarks have slightly
different masses ($m_u\sim5$ MeV and $m_d\sim10$ MeV), the difference
can be ignored in comparison to the scale of the model, $\Lambda\sim1$
GeV.  We observe the second Gegenbauer moments are gradually decrease
as the $n$ increases without sign changes for both of the MIA1 and
MIA2.  On the contrary, the fourth and sixth Gegenbauer moments turn 
negative as the $n$ grows.  It is deeply related to the
fact that the end-point behavior of the pion DAs is changed as the $n$
gets larger.  These sign changes for the fourth and sixth Gegenbauer
moments govern the shapes of the pion DAs around the end-points.  We
also list the results from other theoretical calculations, mainly from
the QCDSR and lattice QCD with specific scales denoted in the
brackets.  Since our scale is about $1$ GeV, in order to compare the present
results with those from Refs.~\cite{Schmedding:1999ap,
RuizArriola:2002bp,Gockeler:2005jz} at the SY scale, we need to evolve
the present results to the corresponding scales.  The anomalous
dimension of the Gegenbauer moments for the one-loop renormalization
equation can be written as follows: 
\bee
\gamma^{(0)}_m=-\frac{8}{3}\left[3+\frac{2}{(m+1)(m+2)}-
4\sum_{m'=1}^{m+1}\frac{1}{m'}\right].
\eee   
We have $\gamma^{(0)}_1=7.11$, $\gamma^{(0)}_2=11.11$,
$\gamma^{(0)}_3=13.95$, $\gamma^{(0)}_4=16.18,\cdots$.  Thus, the
Gegenbauer moments in two different renormalization scales can be
compared by the following relation: 
\bee
a_m(\Lambda_1)=a_m(\Lambda_2)\left[\frac{\alpha(\Lambda_1)}{\alpha(\Lambda_2)}
\right]^{\gamma^{(0)}_m/(2\beta_2)}\simeq 
a_m(\Lambda_2)\left[\frac{\ln[\Lambda_2/\Lambda_0]}{\ln[
\Lambda_1/\Lambda_0]}\right]^{\gamma^{(0)}_m/(2\beta_2)},
\label{evolve}
\eee
where $\beta_0=9$ and $\Lambda_0\simeq200$ MeV.  Using
Eq.~(\ref{evolve}), we can easily compare our results with those at
$2.4$ GeV by multiplying them by $0.72^{\gamma_m^{(0)}/(2\beta_0)}$.
We see that the value $a_2=0.09\sim0.15$ from
Ref.~\cite{Schmedding:1999ap} is quite consistent with the present 
results for the MIA1 ranging $0.07\sim0.13$ incorporated with the
QCD evolution correction, $0.878$. 

\begin{table}[ht]
\begin{center}
\begin{tabular}{c|cccccc}\hline
$n$,MIA&$a^{K}_1$&$a^{K}_2$&$a^{K}_3$&$a^{K}_4$&$a^{K}_5$
&$a^{K}_6$\\   
\hline
1,MIA1&0.06865&0.03264&$-$0.00559&0.00469&0.00315&$-$0.00031\\
2,MIA1&0.08828&$-$0.02634&$-$0.00657&$-$0.02820&0.00022&$-$0.00957\\
3,MIA1&0.09558&$-$0.05145&$-$0.00777&$-$0.03999&$-$0.00214&$-$0.00965\\
\hline
1,MIA2&$-$0.00667&0.03348&0.00160&0.00895&$-$0.00158&0.00088\\
2,MIA2&$-$0.00047&$-$0.01229&0.00010&$-$0.01928&$-$0.00248&$-$0.01017\\
3,MIA2&0.00107&$-$0.03104&$-$0.00083&$-$0.03103&$-$0.00278&$-$0.01120\\
\hline
\cite{Braun:2004vf} (1.0 GeV)&$0.10\pm0.12$&-- &-- &-- &-- &--\\ 
\cite{Khodjamirian:2004ga} (1.0 GeV)&$0.07^{+0.02}_{-0.03}$
&$0.27^{+0.37}_{-0.12}$ &-- &-- &-- &--  \\ 
\cite{Ball:2006wn} (1.0 GeV)&--&$0.30\pm0.15$ &-- &-- &-- &--\\
\cite{Ball:2006fz} (1.0 GeV) &$0.06\pm0.03$ &--&-- &-- &--&--\\
\hline 
\end{tabular}
\end{center}
\caption{The Gegenbauer moments for the kaon distribution
amplitude.  The numbers in the parentheses stand for the scales of
the corresponding works.}
\label{table3}
\end{table}
Now, we consider the kaon Gegenbauer moments. As for the even
Gegenbauer moments, the general tendency is very similar to those of
the pion: The negative sign appears as the $n$ grows, which makes the
DAs suppressed around the end-points.  The results of the second
Gegenbauer moment $a^{K}_2$ are comparable with those from
Refs.~\cite{Ball:2003sc}.   

The most interesting point can be found in the odd Gegenbauer moments
which are finite due to the effects of explicit flavor SU(3)-symmetry
breaking, as shown in Table~\ref{table3}.  As for $a^{K}_1$, we
get all positive values for the MIA1 whereas they are almost
negligible for the MIA2.  As a result, while the kaon DAs turn  
out to be asymmetric for the MIA1, they are almost symmetric for the
MIA2 as shown already in Fig.~\ref{fig1}. Ref.~\cite{Braun:2004vf} provides
the negative value with large uncertainty. 
The positive 
$a^{K}_1$ was suggested in Ref.~\cite{Khodjamirian:2004ga} ranging 
$0.04\sim0.09$ which is about five times larger than those for
the MIA2. When we compare our results with those of
Ref.~\cite{Braun:2004vf,Khodjamirian:2004ga,Ball:2006fz} in which
various techniques in the QCDSR were taken into 
account, the MIA1 gives qualitative agreement with them for 
$a^K_1$ whereas the MIA2 provides much smaller $a^K_1$. The calculated
values of $a^K_2$ are all much smaller than those estimated from the
QCDSR.

In addition to the Gegenbauer moments, we can also define the
expectation value of the momentum which is called $\xi$-moment, as
follows:  
\bee
&&\langle\xi^m\rangle_{\phi}=\int^1_0du\,\xi^m\Phi_{\phi}(u)
\label{ximoment}
\eee
with $\xi=2u-1$.  The numerical results are listed in
Table~\ref{table5}.  We compute the $\xi$-moments upto $m=6$ for pion
and kaon DAs.  As expected from isospin symmetry, the odd $\xi$-moments
for the pion DAs vanish.  As for the kaon DAs, we observe all
negative odd $\xi$-moments for the MIA1 whereas positive for the MIA2, 
which is a similar situation in the case of the Gegenbauer moments.
We also list the $\xi$-moments from other model calculations 
~\cite{DelDebbio:2002mq,RuizArriola:2002bp,Gockeler:2005jz,
Gottlieb:1986ie,Martinelli:1987si,Daniel:1990ah} for comparison.
\begin{table}[ht]
\begin{center}
\begin{tabular}{c|ccc|cccccc} \hline
$n$, MIA&$\langle\xi^2\rangle_{\pi}$&$\langle\xi^4\rangle_{\pi}$&$\langle\xi^6
\rangle_{\pi}$&$\langle\xi^1\rangle_K$&$\langle\xi^2\rangle_K$&
$\langle\xi^3\rangle_K$&$\langle\xi^4\rangle_K$&$\langle\xi^5\rangle_K$&
$\langle\xi^6\rangle_K$\\
\hline
1, MIA1& $0.23776$ &$0.11244$ & $0.06663$ &$0.04119$  & $0.21119$ &$0.01659$&
$0.09366$ &$  0.00902$ &$  0.05326$\\
2, MIA1& $0.21828$ &  $0.09496$ &$  0.05206$ & $0.05297$ &  $0.19097$  & $0.02145$
 & $0.07676$ & $ 0.01149$  & $0.03985$\\
3, MIA1& $0.20995$& $  0.08757$  &$ 0.04606$ &$ 0.05735 $ &$ 0.18236$&  $ 0.02310$
&$ 0.06980$ &$  0.01219 $ &$ 0.03452$\\
\hline
1, MIA2 &$0.23908$ & $ 0.11341$  &$ 0.06734$ &$-0.00400$ &$  0.21148$&$  -0.00141$  
& $0.09430$ &$ -0.00076$  & $0.05394$\\
2, MIA2& $0.22044$  & $0.09641$ & $ 0.05307$ &$-0.00028$  & $0.19579$ & $-0.00010$  
& $0.08090$ &$ -0.00019$  &$ 0.0430$9\\
3, MIA2&$0.21180$ & $ 0.08893$ &$  0.04704$&$  0.00064$ &$  0.18936$ &$  0.00012$  
& $0.07540$ & $-0.00015$ & $ 0.03873$\\
\hline
\cite{DelDebbio:2002mq} (2.67 GeV)&$0.280^{+0.030}_{-0.013}$
&--&--&--&--&--&--&--&--\\
\cite{RuizArriola:2002bp} (2.4 GeV)&$0.040\pm0.005$
&--&--&--&--&--&--&--&--\\
\cite{Gockeler:2005jz} (2.24 GeV)&$0.281(28)$ &--&--&--&--&--&--&--&--\\
\cite{Gottlieb:1986ie} (1.0 GeV)&$1.37\pm0.20$ &--&--&--&--&--&--&--&--\\
\cite{Martinelli:1987si} (1.0 GeV)&$0.25\pm0.10$ &--&--&--&--&--&--&--&--\\
\cite{Daniel:1990ah} (1.0 GeV)&$0.10\pm0.12$
&--&--&--&--&--&--&--&--\\ \hline
\end{tabular}
\caption{$\xi$-moments for the pion and kaon DAs. The numbers in the
  parentheses stand for the scales of the corresponding works.}
\label{table5}
\end{center}
\end{table}   

Now we discuss the inverse moment $I$~\cite{Bakulev:2005cp} and may 
relate to a physical observable in the CLEO 
experiment of $\gamma^*\gamma^*\to \pi^0$ in terms of the pion form
factor~\cite{Gronberg:1997fj}:
\begin{equation}
I=\int^{1}_{0}du\frac{\Phi(u)}{4u(1-u)}.  
\label{cleointegral}
\end{equation}
As shown in Ref.~\cite{Petrov:1997ve}, the value of $I$ turns out to
be exactly $3$ with the asymptotic distribution amplitude.  In
Table~\ref{table6}, we list the results of $I$ for the MIA1 and MIA2
with the $n$ varied.  Although the results are similar to that of the
asymptotic one ($I=3$), they in fact deviate from it by approximately
$10\%$.  Note that $I_{\pi}$ and $I_K$ are very similar each other
($I_K\lesssim I_{\pi}$). 
\begin{table}[ht]   
\begin{tabular}{c|ccc|ccc} \hline
& \multicolumn{3}{c|}{$I_{\pi}$} & \multicolumn{3}{c}{$I_K$} \\ 
 \hline
$n$ & $1$ & $2$ & $3$  & $1$ & $2$ & $3$ \\ 
\hline
MIA1&3.40471&3.00196&2.88593&3.12749&2.79872&2.70349\\
MIA2&3.42072&3.02038&2.90240&3.14755&2.85608&2.77235\\ \hline
\end{tabular}
\caption{The results of the integral $I$.}
\label{table6}
\end{table}

We now discuss the expectation value of the transverse
momentum $\langle k^{m}_T\rangle$.  As indicated in
Refs.~\cite{Chernyak:1977fk,Praszalowicz:2001pi}, it contains
important information on the meson in hard processes.  Moreover, it is
deeply related to the QCD condensates~\cite{Zhitnitsky:1993vb}.  The
moments of $\Phi(k_T)$ can be written as follows:    
\begin{eqnarray}
\langle k^{m}_T\rangle=\frac{1}{iF_{\phi}P\cdot n}\langle0|\bar{q}_f
\rlap{/}{n}\gamma_5[it_{\mu}\tilde{D}^{\mu}]^{m}q_g|
\phi(P)\rangle,
\end{eqnarray}
where $\tilde{D}^{\mu}=\overrightarrow{D}^{\mu}-\overleftarrow{D}^{\mu}$ is
the covariant derivative in which $\overrightarrow{D}^{\mu}=\overrightarrow{
\partial}+ig A_{\mu}\cdot \lambda/2$, and $t_{\mu}$ stands for the unit
vector in the transverse direction.  We note that the $k_T$ dependence of the
meson light-cone wave function $\Psi(u,k_T)$ is derived from the quark 
propagator ($k^2=k_+k_--k^2_T$) as shown in Eq.~(\ref{DAAV}).  In Refs.~\cite{
Chernyak:1977fk,Zhitnitsky:1993vb} the $\langle k^{m}_T\rangle$ is
approximated in terms of quark and quark-gluon mixed condensates with
the soft pion theorem in the chiral limit: 
\begin{eqnarray}
\langle k^{2}_T\rangle=\frac{5}{36}\frac{\langle ig\bar{q}
\sigma_{\mu\nu} G^{\mu\nu}q\rangle}{\langle\bar{q} q
\rangle}.  
\label{k24}
\end{eqnarray}
Hence, there are two different ways to compute $\langle
k^{m}_T\rangle$ in the chiral limit: Firstly, it can be computed by
using Eq.~(\ref{k24}) as done in Ref.~\cite{Nam:2006ng} within the
same scheme as the present work.  Secondly, it can be directly
calculated as follows: 
\begin{eqnarray}
\langle k^{m}_T\rangle=\int^{\infty}_0 d^2k_T\,\, k^{m}_T{\Phi}
(k_T)=\int^1_{0}du\int^{\infty}_0 d^2k_T\,\, k^{m}_T\Psi(k_T,u).
\label{ktintegral}
\end{eqnarray}
\begin{figure}[ht]
\begin{tabular}{cc}
\resizebox{8cm}{6cm}{\includegraphics{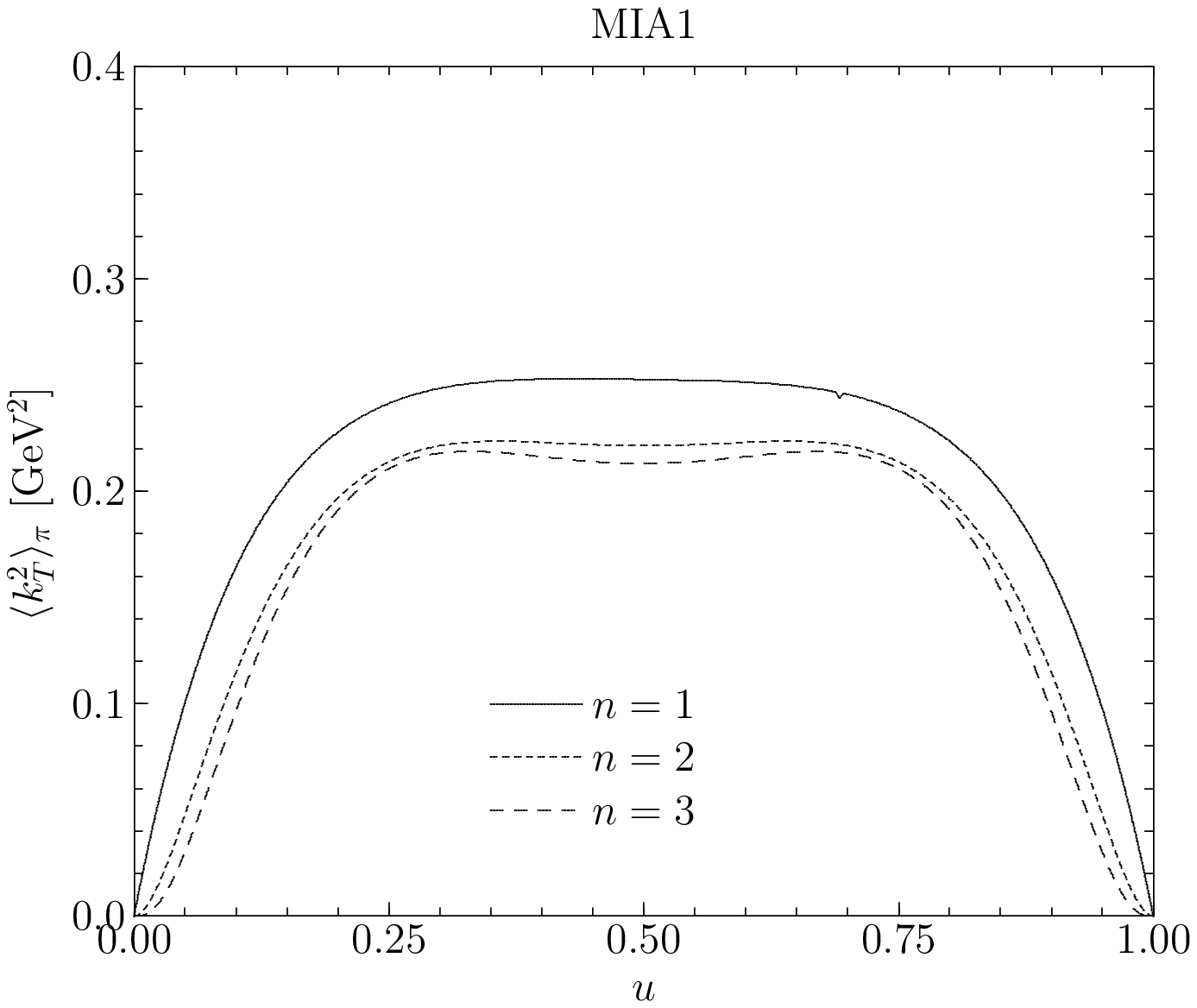}} 
\resizebox{8cm}{6cm}{\includegraphics{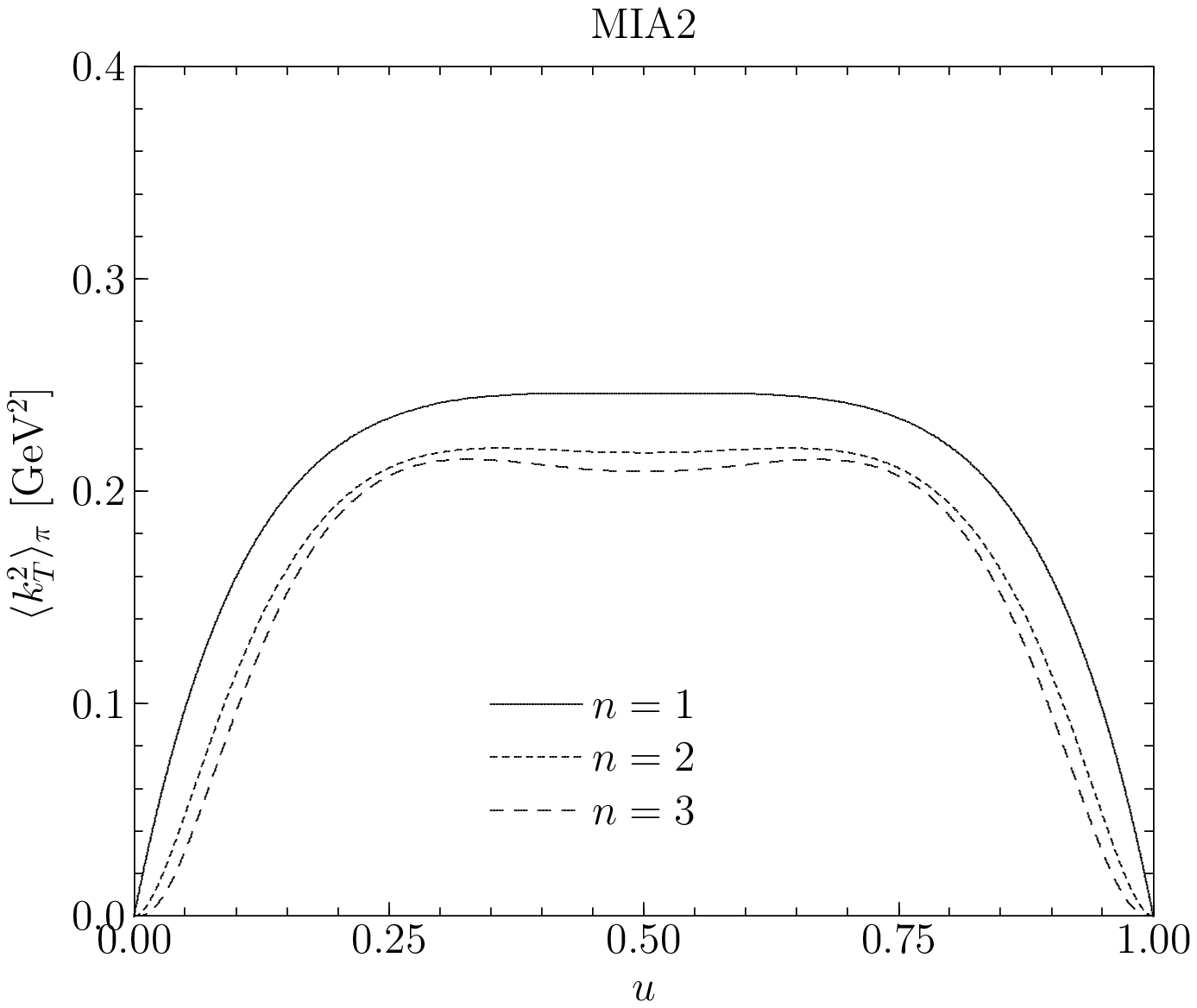}} 
\end{tabular}
\begin{tabular}{cc}
\resizebox{8cm}{6cm}{\includegraphics{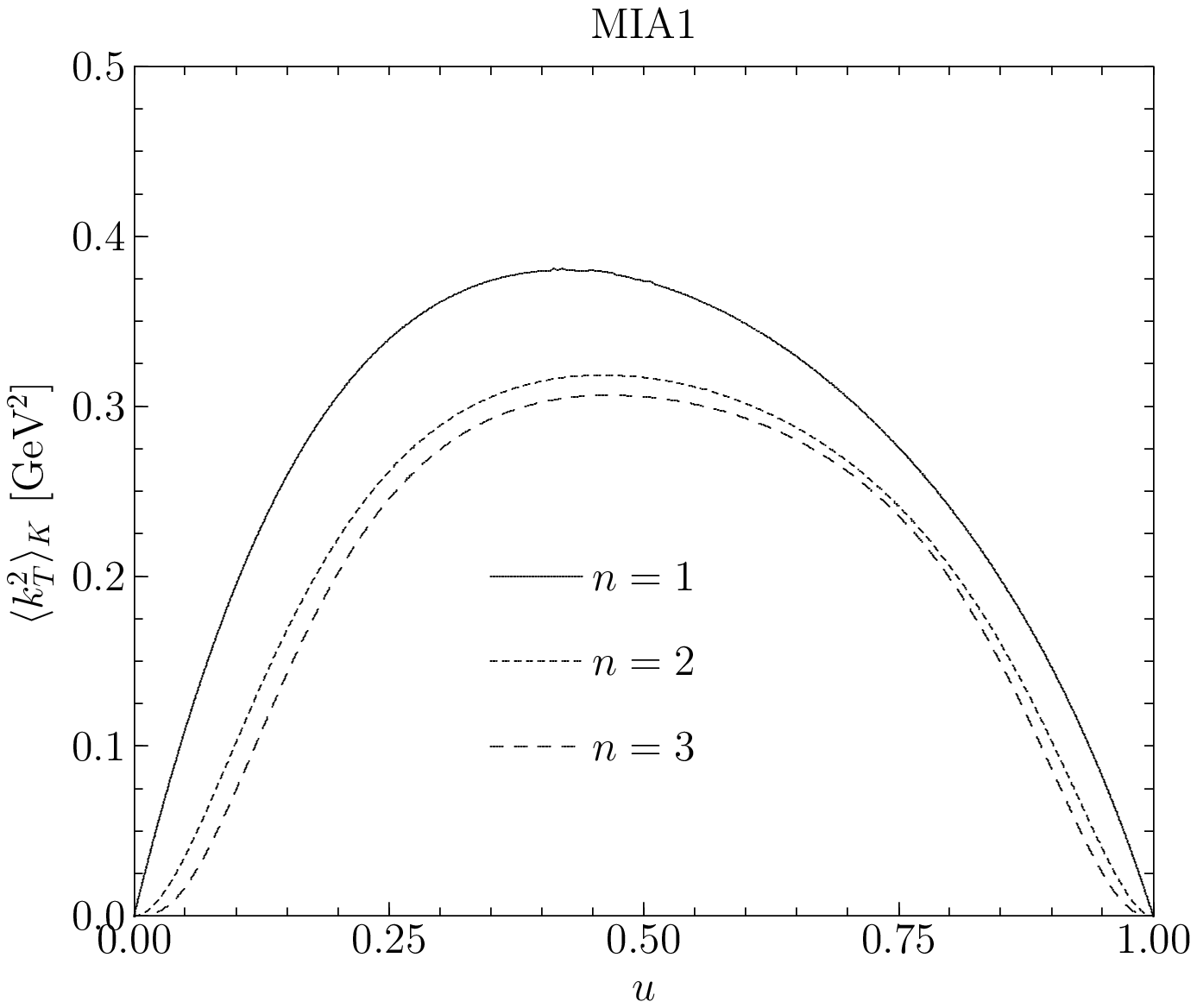}} 
\resizebox{8cm}{6cm}{\includegraphics{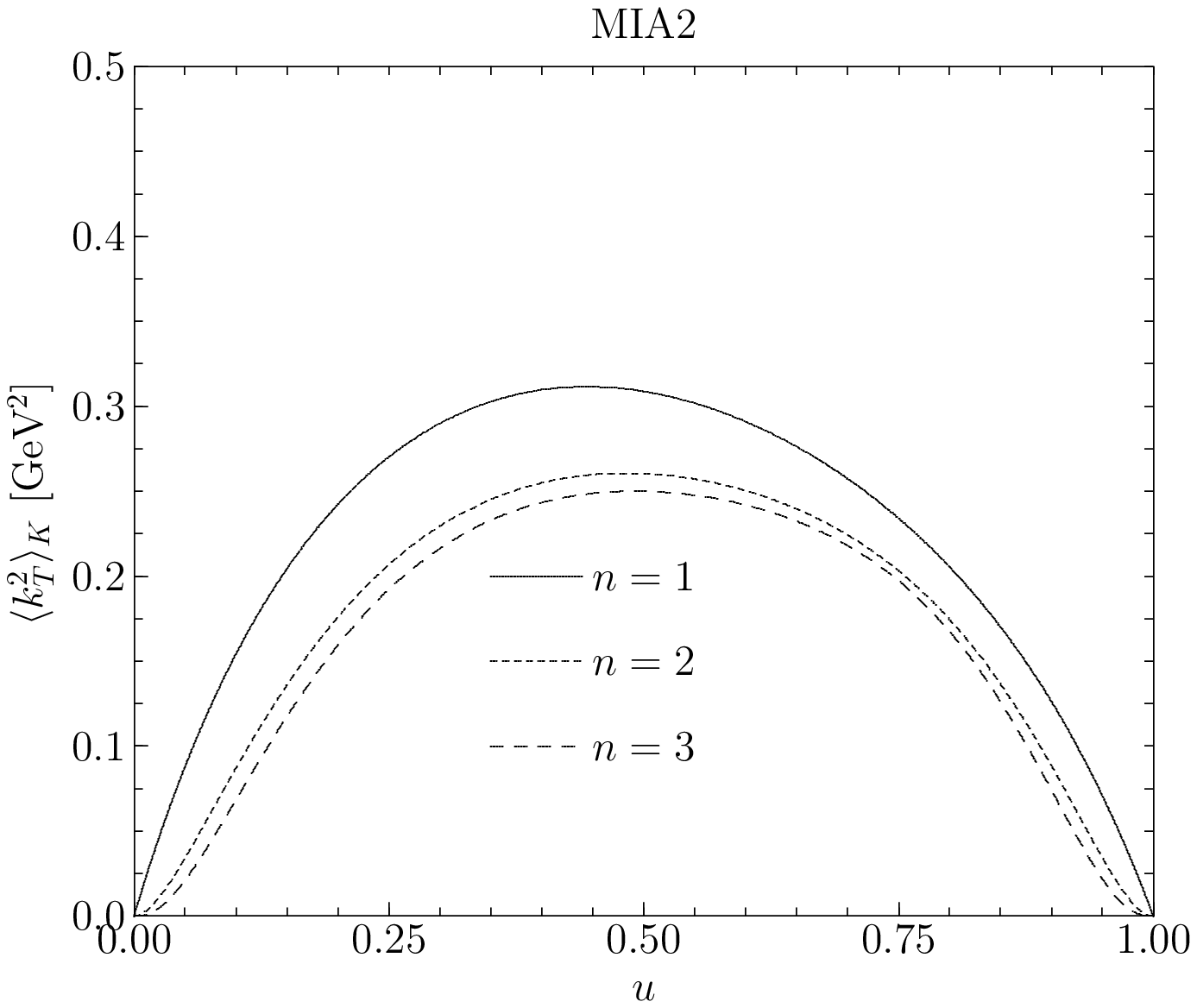}} 
\end{tabular}
\caption{$\langle k^2_T\rangle$ for the pion (upper two
panels) and kaon (lower two panels) for different $n$.}
\label{fig3}
\end{figure}
Since the light-cone wave function $\Psi (k_T,u)$ can be obtained by
integrating Eq.~(\ref{DAAV}) over $k^+$ and $k^-$, it is possible to
calculate $\langle k^{m}_T\rangle$ given in Eq.~(\ref{ktintegral}).
In the present work, we only consider the cases $m=2$ and $4$.  In
Fig.~\ref{fig3}, we depict the numerical results for $\langle
k^2_T\rangle_{\pi,K}$.  As for the pion, the transverse
momentum is rather equally distributed between the 
quark and the anti-quark inside it, so that  $\langle
k^2_T\rangle_{\pi}$ is almost flat within the range
$0.25\lesssim u\lesssim0.75$.  On the contrary, $\langle
k^2_T\rangle_K$ shows asymmetry, since the light quark and 
the strange quark inside the kaon carry different fractions of the 
transverse momentum due to their mass difference.  
\begin{figure}[ht]
\begin{tabular}{cc}
\resizebox{8cm}{6cm}{\includegraphics{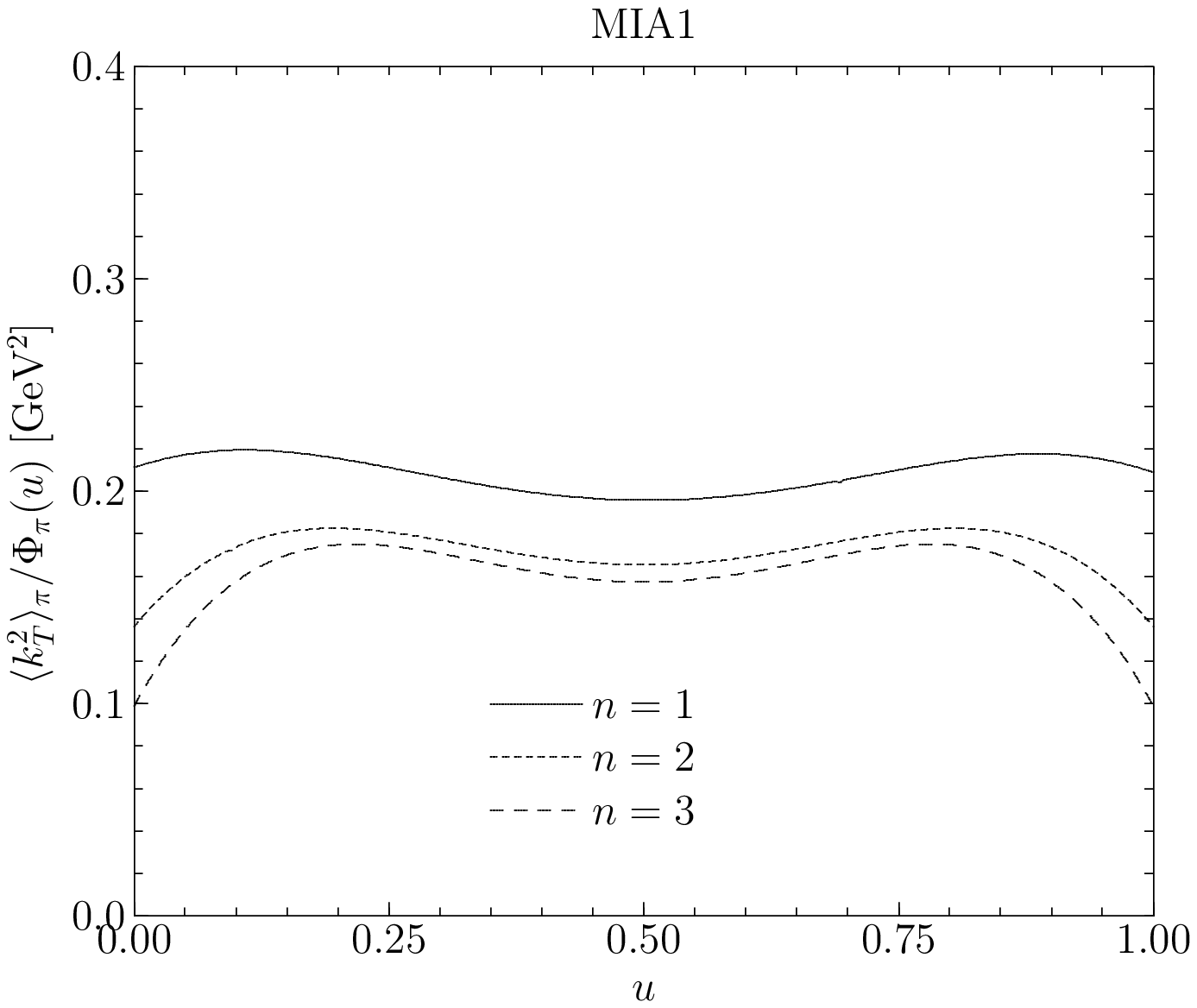}} 
\resizebox{8cm}{6cm}{\includegraphics{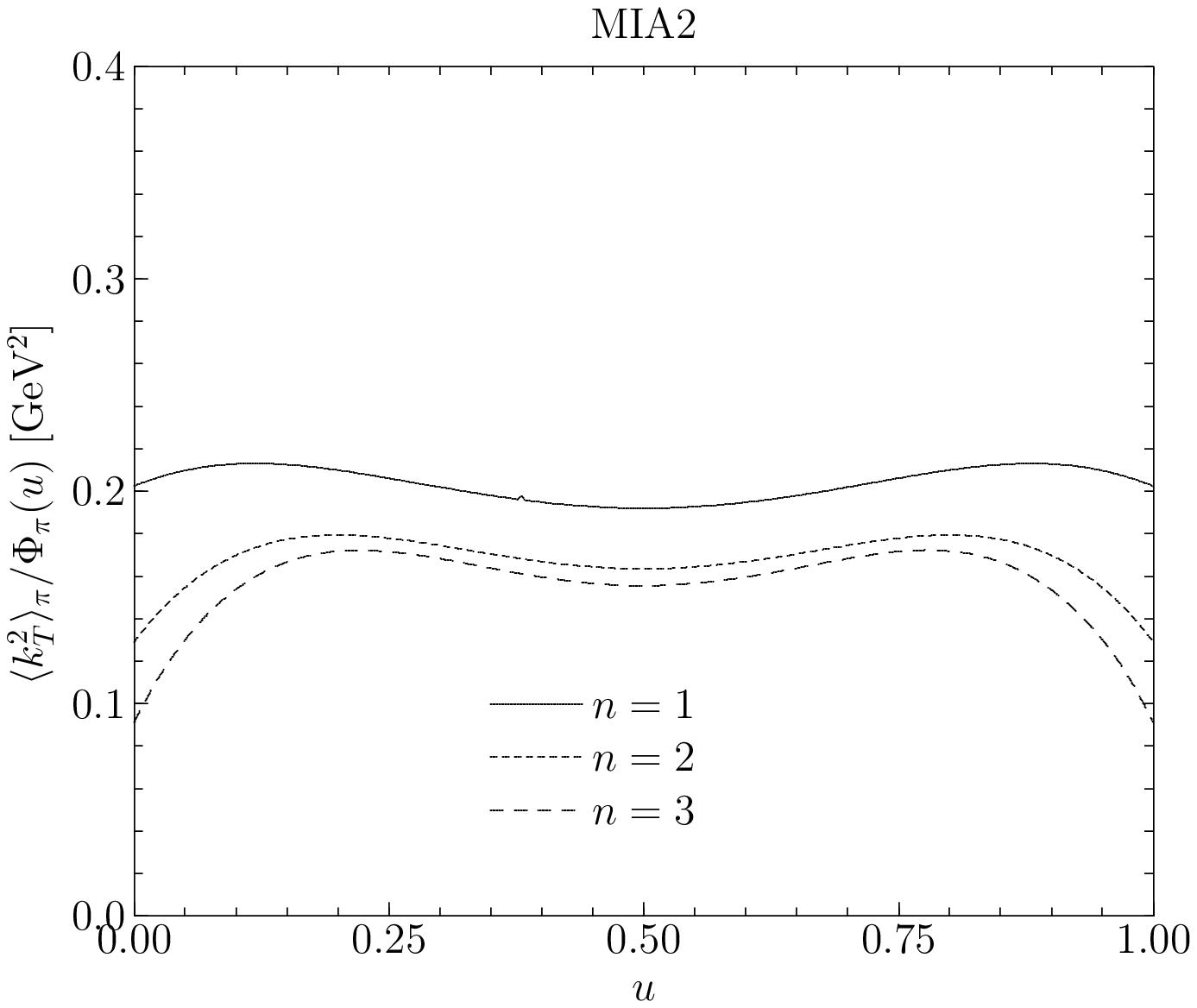}}  
\end{tabular}
\begin{tabular}{cc}
\resizebox{8cm}{6cm}{\includegraphics{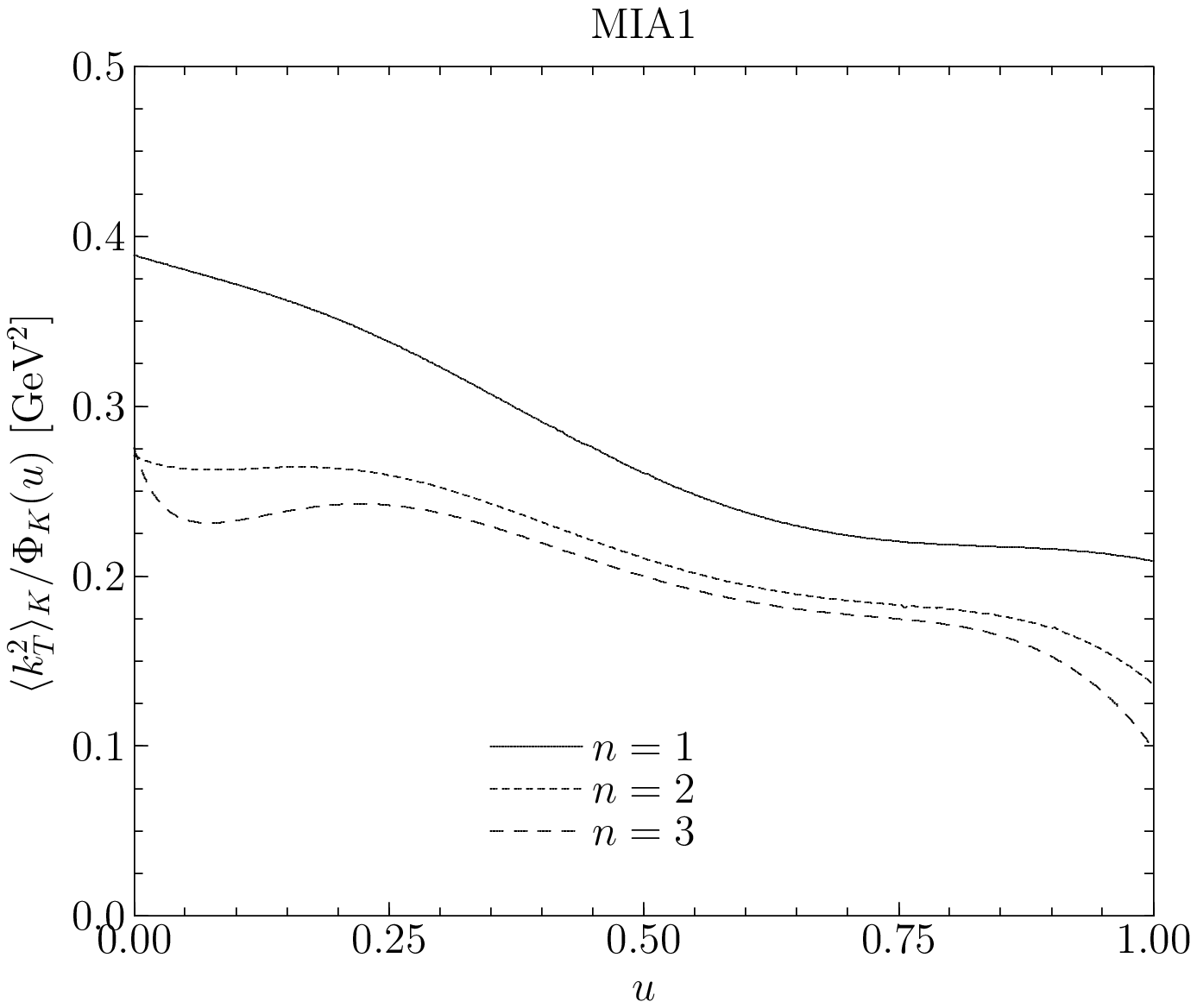}} 
\resizebox{8cm}{6cm}{\includegraphics{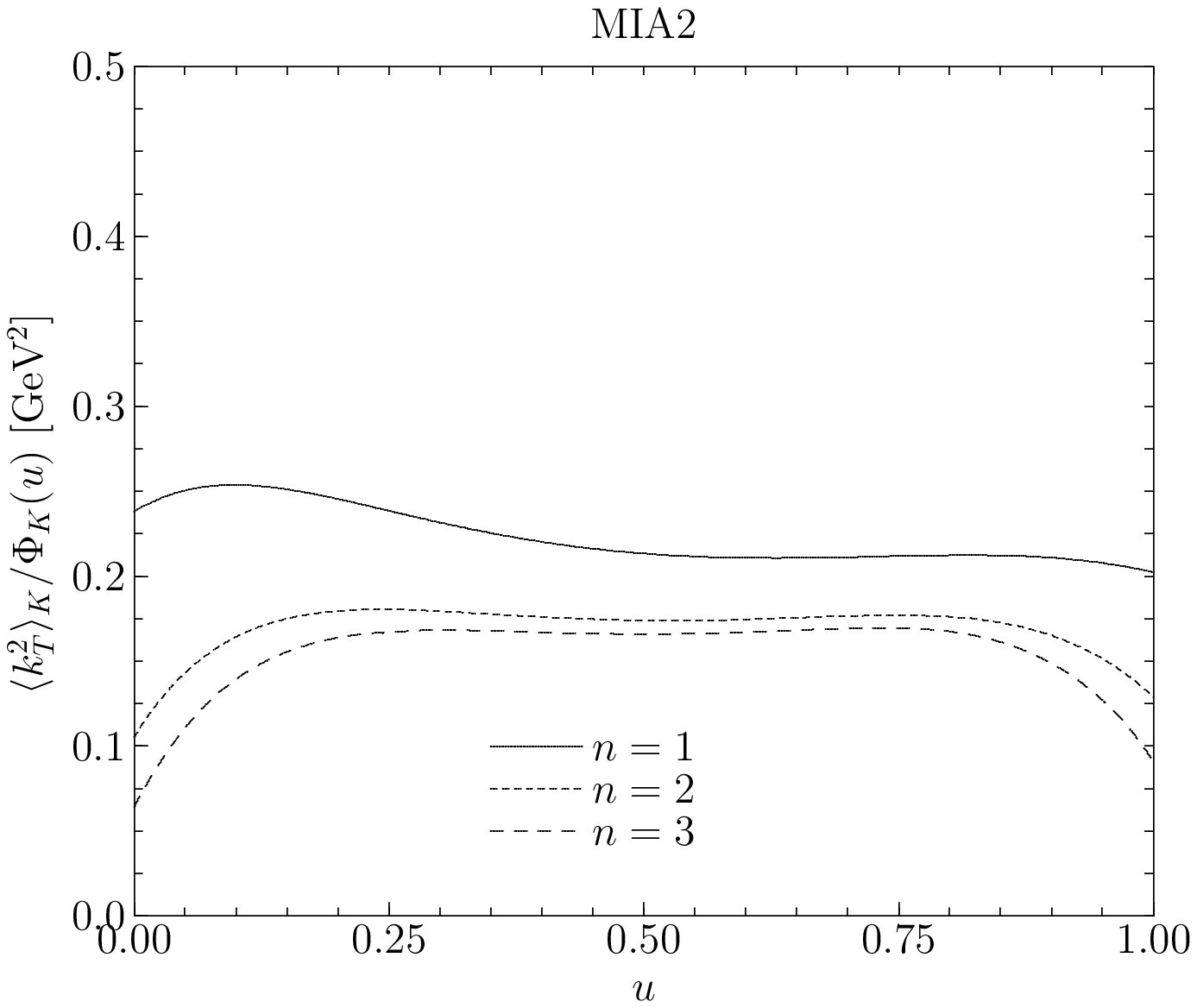}}  
\end{tabular}
\caption{$\langle k^2_T\rangle/\Phi$ for the pion (upper two
  panels) and kaon (lower two panels) for different $n$.} 
\label{fig4}
\end{figure}
Using the results of $\langle k^2_T(u)\rangle$, we test now 
the factorization hypothesis of the light-cone wave functions for the
pion and kaon: The light-cone wave function can be factorized into the 
longitudinal and transverse contributions in a separable form
$\Psi(u,k_T)=\Phi(u)\Phi(k_T)$.  This hypothesis is useful 
when the meson light-cone wave function is to be integrated over the quark
four- momentum.  However, as discussed in
Ref.~\cite{Zhitnitsky:1993vb}, the hypothesis does not work well.
It can be understood in the following equation:   
\begin{eqnarray}
\frac{\langle k^2_T(u)\rangle}{\Phi(u)}=\frac{\int d^2k_T\,\,
k^{2m}_T\Psi(k_T,u)}{\int d^2k_T\,\,\Psi(k_T,u)}\to\frac{\int
d^2k_T\,\, k^{2m}_T\Phi(u){\Phi}^{}(k_T)}{\int d^2k_T\,\,
\Phi(u)\Phi(k_T)}=\frac{\int d^2k_T\,\, k^{2m}_T\Phi(k_T)}{\int d^2k_T\,\,
\Phi(k_T)}.\nn\\  
\label{factorization}
\end{eqnarray}
Thus, if the factorization hypothesis had worked well, then the
quantity $\langle k^2_T(u)\rangle/\Phi(u)$ would have been independent of
$u$.  In the upper two panels of Fig.~\ref{fig3}, we show the
numerical results of $\langle k^2_T(u)\rangle/\Phi(u)$ for the
pion.  As shown in Fig.~\ref{fig4}, the results are rather flat for the
region $0.25\lesssim u\lesssim0.75$ but not constant, which is
similar to those of Ref.~\cite{Praszalowicz:2001pi}.  
In the kaon case, the factorization hypothesis is obviously a wrong
one.  As shown in the lower panel of Fig.~\ref{fig4}, the results
are not at all constant.  

Finally, we calculate the averaged expectation values of the
transverse momenta with respect to $u$ so that we can compare them
with those obtained above, using the 
following equation: 
\begin{eqnarray}
\langle k^m_T\rangle_{\mathrm{Averaged}}=\int^1_{0}du\,dk^2_T
\,k_T\Psi(u,k_T). 
\end{eqnarray}
In Table~\ref{table7}, we list the results of $\langle k^2_T\rangle$,
$\langle k^4_T\rangle$ and $R=\langle k^4_T\rangle/\langle
k^2_T\rangle^2$.  Interestingly, we get rather stable results of the 
expectation values for $n=2,3$.  However, we obtain a noticeably
large result for $n=1$ due to the larger values of $\langle
k^4_T\rangle$ for $n=1$.  As for the kaon, larger values of $\langle
k^4_T\rangle$ are yielded, in particular, $R\sim20$ with
$n=1$ for the MIA2.  We note that the ratio $R$ was estimated to
be $5\sim 9$ for the pion DA in
Refs.~\cite{Zhitnitsky:1993vb,Chernyak:1981zz}, which is similar to
the present results, especially, in the case of $n=1$ ($\sim 5$).  
However, in general, our results turn out to be smaller and less broad 
than those from other models.  We also list those 
for $\langle k^2_T\rangle_{\pi}$ from
Ref.~\cite{RuizArriola:2002bp,Zhitnitsky:1993vb}.  The result of  
Ref.~\cite{Zhitnitsky:1993vb} is smaller than ours.  
\begin{table}[ht]
\begin{tabular}{c|ccc|ccc} \hline
$n$,MIA&$\langle k^2_T\rangle_{\pi}$&$\langle
k^4_T\rangle_{\pi}$&$\langle k^4_T\rangle_{\pi}/\langle
k^2_T\rangle_{\pi}^2$&$\langle k^2_T\rangle_K$&$\langle 
k^4_T\rangle_K$ &$\langle k^4_T\rangle_K/\langle
k^2_T\rangle_K^2$\\
\hline
1,MIA1& 0.20671 &  0.22063 &  5.16348& 0.26937 &  1.12985 & 15.57098\\
2,MIA1&0.17382  & 0.08618 &  2.85241&  0.21271&   0.15226 &  3.36524\\
3,MIA1& 0.16554 &  0.07322  & 2.67164& 0.19980 &  0.11746 &  2.94253 \\
\hline
1,MIA2& 0.20230 &  0.21882 &  5.34684 & 0.22160 &  0.96318 & 19.61359\\
2,MIA2&0.17156  & 0.08461 &  2.87481& 0.17503 &  0.12040 &  3.93006\\
3,MIA2& 0.16302 &  0.07166 &  2.69623 &0.16391 &  0.09156 &  3.40782\\
\hline
\cite{RuizArriola:2002bp} ($1$ GeV) &0.18490&--&--&--&--&--\\
\cite{Zhitnitsky:1993vb} ($1$ GeV) &0.10000&--&--&--&--&--\\ \hline
\end{tabular}
\caption{$\langle k^2_T\rangle$ [GeV$^2$],  $\langle
  k^4_T\rangle$ [GeV$^4$] and $\langle k^2_T\rangle/\langle
  k^2_T\rangle^2$ for the pion and kaon DAs for different
  $n$ and MIA.}
\label{table7}
\end{table}

\section{Summary and Conclusion}
In the present work, we aimed at investigating the leading-twist
light-cone distribution amplitudes of the pion and kaon, the effects
of flavor SU(3)-symmetry breaking being explicitly considered from the
QCD instanton vacuum. 

We started from the modified improved effective chiral action from the
instanton vacuum, with two different ways considered: The dynamical
quark mass without the current-quark mass correction factor (MIA1) and
that with it (MIA2).  As for the momentum-dependent dynamical
quark mass, which plays an essential role in describing the meson
distribution amplitudes, we parameterized it by a simple-pole type
form factor with three different powers, $n=1,2,3$.  The cut-off mass
was fixed to reproduce pion and kaon decay constants simultaneously in
such a way that the normalization condition of the distribution
amplitudes may be satisfied.  

We first examined the pion and kaon distribution amplitudes in detail.
As expected from the light current-quark masses ($m_{u,d}=5$ MeV), we
did not find any noticeable difference between the pion distribution
amplitudes for the MIA1 and those for the MIA2.  It was found that the
pion distribution amplitudes turn out to be symmetric for the momentum
fraction $u$, while the kaon ones show asymmetric behavior due to flavor
SU(3)-symmetry breaking effects.  However, when the 
current-quark mass dependence of the dynamical quark mass is
considered, the kaon distribution amplitudes become less asymmetric. 
Furthermore, we found that as the power $n$ increases, the 
shape of the distribution amplitudes is getting flat.  It was also
shown that the end-point behavior of the distribution amplitudes
varies as the power of the form factors is changed.   

In order to see how far the pion and kaon distribution amplitudes
deviate from the asymptotic one and how large the flavor SU(3)-symmetry
breaking effects are, we calculated the Gegenbauer moments of 
the distribution amplitudes to the sixth order.  The odd Gegenbauer
moments turn out to vanish for the pion distribution amplitudes due to
isospin symmetry, whereas they remain finite for the kaon one due to
the effects of explicit flavor SU(3)-symmetry breaking.  The fourth
and sixth Gegenbauer moments turn negative as the power $n$ 
increases.  It implies that the pion distribution amplitude at the 
end-points ($u=0,1$) is changed to be concave, while it becomes flat
in the vicinity of $u=0.5$, as mentioned above.
As for the kaon Gegenbauer moments, we observed very similar tendency
to those of the pion, except for the odd Gegenbauer moments. The 
current-quark mass corrections to the dynamical quark mass made the
first Gegenbauer moment nearly negligible.  

We also took into account the QCD evolution via the QCD renormalization 
group equation in order to compare the present results for the
Gegenbauer moments with other empirical or theoretical estimates at
the Schmedding-Yakovlev scale ($2.4$ GeV). 

Finally, the expectation values of the transverse momentum $\langle
k^{2m}_T\rangle$ were computed by using the distribution amplitudes
and light-cone wave functions $\Psi(k_T,u)$.  As expected from
explicit flavor SU(3)-symmetry breaking, the quarks inside the kaon 
carry the momentum fractions in an asymmetric manner, 
whereas those inside the pion do symmetrically.  Using these results,
we tried to analyze the factorization hypothesis of the total
light-cone wave function, $\Psi(u,k_T)=\Phi(u)\Phi(k_T)$.  The
numerical results of this ratio for the pion look rather flat but not
constant.   Those for the kaon are far from the flat shape.  Thus,
altogether, the factorization hypothesis is invalid for the meson
light-cone wave functions.  We also calculated the momentum
fluctuation shown in the ratio of $\langle 
k^4_T\rangle/\langle k^2_T\rangle^2$ and compared it with other 
model calculations.

\section*{Acknowledgment}
The present work is supported by the Korean Research Foundation
(KRF--2003--070--C00015).  The authors appreciate the critical and
constructive comments from N.~G.~Stefanis, M.~Prasza\l
owicz, A.~Dorokhov, R.~Zwicky and A.~Stolz. S.I.N. is grateful to  
Y.~Kwon for fruitful discussions. S.I.N. also appreciate the
hospitality of the Research Center for Nuclear Physics (RCNP) in Japan
where the part of this work was done. The work of A.H. was
supported in part by the Grant for Scientific Research ((C)
No.16540252) from the Ministry of Education, Culture, Science and
Technology of Japan.

\section*{Appendix}
The definition and usage of light-cone coordinate vectors. 
\bee
\hat{n}_{\mu}&=&(1,0,0,1),\,\hat{n}'_{\mu}=(1,0,0,-1),\,n\cdot
\bar{n}=2,\nn\\ 
k_{\mu}&=&\frac{k^{+}}{2}\hat{n}_{\mu}+\frac{k^{-}}{2}\hat{n}'_{\mu}+k_{T\mu},\nn\\ 
k_{T\mu}&=&(0,k_1,k_2,0),\,k^{\mu}_{T}=(0,-k_1,-k_2,0),\nn\\
k_{\mu}&=&\frac{k_0+k_3}{2}(1,0,0,1)+\frac{k_0-k_3}{2}(1,0,0,-1)+(0,k_1,k_2,0),\nn\\k^{\mu}&=&\frac{k_0-k_3}{2}(1,0,0,-1)+\frac{k_0+k_3}{2}(1,0,0,1)+(0,-k_1,-k_2,0),\nn\\k^{+}&=&k_0+k_3=k\cdot \hat{n},\,k^{-}=k_0-k_3=k\cdot\hat{n}',\,k\cdot k=k^{+}k^{-}-k_{T}^2,\nn\\d^4k&=&\frac{1}{2}dk^+dk^-dk^2_{T}.
\label{light_cone_vectors}
\eee





\end{document}